\documentclass[12pt]{article}
\usepackage{hyperref}
\usepackage[dvips]{graphicx}
\renewcommand{\baselinestretch}{1.5}
\def\be{\begin{equation}}
\def\ee{\end{equation}}
\def\beq{\begin{eqnarray}}
\def\eeq{\end{eqnarray}}

\begin{document}

\vspace{.5cm}

\begin{center}
{\bf \large Statistics of Neutron Stars at the Stage of Supersonic Propeller} 
\footnote{Astronomy Letters, 2005, {\bf 31}, (9), 579-585, Translated from Russian by V. Astakhov}
\end{center}

\vspace{1cm}

\begin{center}
V. S.~Beskin$^1$ and S. A.~Eliseeva$^2$ \\
\vspace{1cm}
$^1${\it P.N.~Lebedev Physical Institute, Leninsky prosp. 53, Moscow, 119991, Russia \\
$^2$Moscow Institute of Physics and Technology, Dolgoprudny, Moscow region, 141700, Russia}
\end{center}

\begin{center}
{$\rm ^1 beskin@lpi.ru$, $\rm ^2sa\_eliseeva@comcast.net$}
\end{center}

\vspace{1cm}

\begin{center}
{\bf Abstract}
\end{center}

{\small 
We analyze the statistical distribution of neutron stars at the stage of a supersonic propeller. An important point of our analysis is allowance for the evolution of the angle of inclination of the magnetic axis to the spin axis of the neutron star for the boundary of the transition to the supersonic propeller stage for two models: the model with hindered particle escape from the stellar surface and the model with free particle escape. As a result, we have shown that a consistent allowance for the evolution of the inclination angle in the region of extinct radio pulsars for the two models leads to an increase in the total number of neutron stars at the supersonic propeller stage. This increase stems from he fact that when allowing for the evolution of the inclination angle $\chi$ for neutron stars in the region of extinct radio pulsars and, hence, for the boundary of the transition to the propeller stage, this transition is possible at shorter spin periods ($P \sim 5-10$ s) than assumed in the standard model.}

\vspace{1cm}
 
\section{Introduction}

Previously, we showed \cite{BE2005} that the transition to the propeller stage could occur at much shorter periods than assumed previously when allowance is made for the evolution of the angle of inclination of the magnetic axis to the spin axis of the neutron star. Consequently, allowance for the evolution of the inclination angle at the stage of extinct radio pulsars would affect significantly the statistics of neuron stars at the propeller stage.

As \cite{DFP1979} and \cite{DP1981} showed, the following two so-called substages that a neutron star passes as its spin period decreases must be distinguished: supersonic and subsonic propellers. In both cases, the neutron star spins down due to the interaction of the stellar magnetosphere with the surrounding matter.

As a neutron star spins down, the electrodynamic processes cease to play a significant role in the pulsar's magnetosphere. When the radius at which the pressure of the ambient medium is balanced by the magnetodipole radiation pressure (the Schwarzman radius) becomes equal to the Alfven radius, the star passes to the stage of a supersonic propeller. As the spin period increases further, the barrier produced by the rotation of the neutron star?s strong magnetic field ceases to be efficient, and the plasma begins to penetrate into the pulsar?s magnetosphere. Thus, the pulsar passes to the stage of a subsonic propeller.

Subsequently, due to an even larger spindown, the accretion rate increases and can reach its maximum at a certain period $P$ -- the neutron star passes to the stage of a steady accretor. The evolution of a neutron star at the "supersonic propeller--subsonic propeller--steady accretor" stages was considered in more detail in the paper by \cite{I2003}, whose results we used here.

In this paper, we consider the stationary distribution of isolated neutron stars at the supersonic propeller stage. As previously \cite{BE2005}, all our calculations were performed for two basic models: the model with hindered particle escape from the neutron-star surface (\cite{RS1975}, \cite{BGI1993}) and the model with free particle escape (\cite{A1979}, \cite{M1999}). We disregard the possibility of magnetic-field evolution and assume that no neutron stars are born at the supersonic propeller stage. An important point of our study is that the boundary of the transition of a neutron star from the ejector stage to the propeller stage is presented with allowance for the evolution of the inclination angle.

As a result, we obtained the distribution of neutron stars at the supersonic propeller stage in spin period $P$. We show that the number of neutron stars at this stage is larger when the evolution of the inclination angle is taken into account for the transition boundary than that in the case where the standard model is considered. We also show once again that when the evolution of the inclination angle $\chi$ is taken into account, neutron stars can pass to the propeller stage even at fairly short periods: $P \sim 5-10 \mbox{ s}$.

\section{Basic equations}

Lipunov and Popov \cite{LP1995} formulated an important assertion: for a constant magnetic field, the ejector stage at reasonable parameters is always longer than the propeller stage. Our objective in this part of the paper is to determine the number of neutron stars at the propeller stage disregarding both the magnetic-field decay and the evolution of the angle of inclination of the magnetic axis to the spin axis of the star at this stage.

A neutron star at the propeller stage is known to spin down due to the transfer of angular momentum to the surrounding matter (\cite{S1970}, \cite{IS1975}). There are many formulas that describe this spindown (see \cite{L1987}, \cite{LP1995}). Nevertheless, virtually all of them can be reduced to the form \cite{PP2002}
\begin{equation}
\frac{dI \omega}{dt}=-k_t \frac{\mu^2}{R_A^3}.
\end{equation}

\noindent Here $I \sim 2/5 MR^2$ is the moment of inertia of the star, $\mu$ is the magnetic dipole moment, $k_t \approx 1$ is a dimensionless constant, and $R_A$ is the Alfven radius:
\begin{equation}
R_A=\left( \frac{\mu^2}{\dot M \sqrt{2GM}}\right)^{2/7},
\end{equation}

\noindent where $M$ is the mass of the neutron star. The acceleration rate can be estimated as
\begin{equation}
\dot M \simeq \pi R_{\alpha}^2 \rho V_{\infty},
\end{equation}

\noindent where $V_{\infty}$ is the space velocity of the neutron star relative to the surrounding interstellar medium, $\rho$ is the density of the interstellar medium, and $R_{\alpha}$ is the gravitational capture radius, which for spherically symmetric accretion is
\begin{equation}
R_{\alpha}=\frac{2GM}{V_{\infty}^2}.
\end{equation}

The maximum possible accretion rate onto an isolated neutron star (for spherically symmetric accretion) was determined by Ikhsanov \cite{I2003} and can be specified by the following expression:
\begin{equation}
\dot M \leq M_{max} \simeq 2.3 \cdot 10^{11}\rho_{-24}m^2V_6^{-3} \rm gs^{-1},
\end{equation}

\noindent where $m$ is the mass of the neutron star in terms of solar masses, $\rho_{-24}=\rho/10^{-24} \mbox{ g cm}^{-3}$, and $V_6=V_{\infty}/10^6 \mbox{ cm s}^{-1}$. Given all of the factors listed above, the transfer of angular momentum to the surrounding interstellar medium is found to cause the neutron star to spin down at the rate
\begin{equation}
\frac{dP}{dt}=7.3 \cdot 10^{-14}B_{12}^{2/7}P^2.
\label{PR_dP_dt}
\end{equation}

\renewcommand{\baselinestretch}{1.0}

\begin{figure}
\centering
\includegraphics[width=10cm, height=10cm, keepaspectratio]{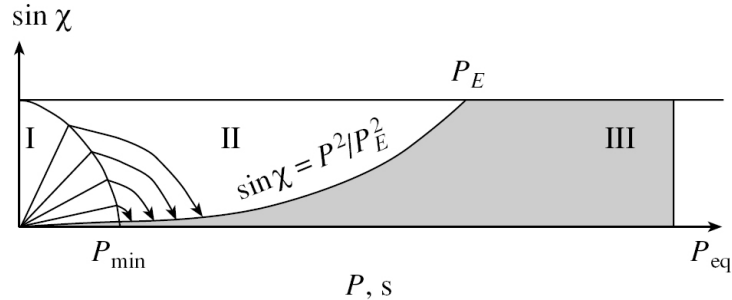}
\caption{\small{Evolution of neutron stars under the assumption of hindered particle escape: I -- active radio pulsars, II -- extinct radio pulsars, and III -- pulsars at the supersonic propeller stage.}}
\label{fig1}
\end{figure}

\renewcommand{\baselinestretch}{1.5}
Further, it should be noted that for the model with hindered particle escape from the pulsar's surface, the transition to the supersonic propeller stage for most neutron stars can occur only at fairly small angles $\chi$ between the magnetic axis and the spin axis (Fig.~\ref{fig1}). Consequently, in the case of hindered particle escape, the neutron stars passing to the propeller stage are almost aligned rotators, and the subsequent evolution of the inclination angle is unimportant. As a result, it can be assumed with a sufficient accuracy for the Ruderman-Sutherland model that the inclination angle does not evolve at the supersonic propeller stage.

\renewcommand{\baselinestretch}{1.0}

\begin{figure}
\centering
\includegraphics[width=10cm, height=10cm, keepaspectratio]{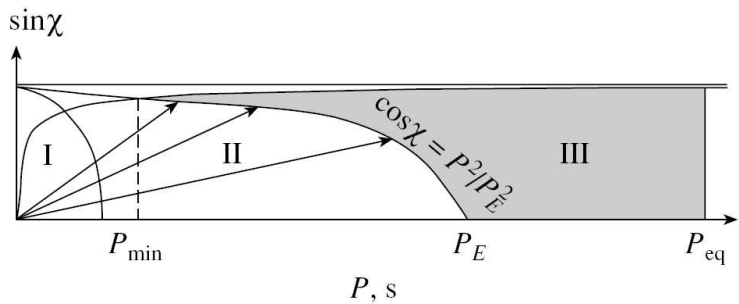}
\caption{\small{Evolution of neutron stars under the assumption of free particle escape. The notation is the same as that in Fig.~\ref{fig1}.}}
\label{fig2}
\end{figure}

\renewcommand{\baselinestretch}{1.5}

As regards the model with free particle escape from the neutron-star surface, the characteristic period at which the transition to the propeller stage occurs is much longer than the death period. Consequently, most of the neutron stars will pass to the region of angles close to $90^{\circ}$ before the transition of the star to the supersonic propeller stage (see Fig.~\ref{fig2}).

Thus, we are interested in the statistics of pulsars at the supersonic propeller stage without considering the angle of axial inclination. Accordingly, to determine the stationary distribution function of neutron stars, let us introduce their distribution $N_3(P,B)$ in period $P$ and magnetic field $B$. Next, as in the case of extinct radio pulsars, we assume that the function $N_3(P,B)$ depends on the magnetic field $B$ of the neutron star only as a parameter; i.e., we disregard the magnetic-field evolution over the lifetime of the neutron star. As a result, the kinetic equation for neutron stars at the supersonic propeller stage can be written as
\begin{equation}
\frac{\partial}{\partial P} \left(N_3 \frac{dP}{dt} \right)=F(P,B),
\label{PR_kin}
\end{equation}

\noindent where the source function $F(P,B)$ is determined by the neutron-star flux from the region of extinct radio pulsars.

Finally, we must specify the boundary periods for the supersonic propeller stage, i.e., the period at which the transition from the ejector stage to the propeller stage occurs and the period at which the accretion of matter from the interstellar medium onto the neutron star begins. As we found previously \cite{BE2005}, the transition from the ejector stage to the propeller stage occurs for the model with hindered particle escape from the neutron-star surface at the spin period
\begin{equation}
P_{pr}=P_E sin^{1/2}\chi.
\label{PR_PPR_hind}
\end{equation}

At the same time, the following expression holds for the model with free particle escape: 
\begin{equation}
P_{pr}=P_E cos^{1/2}\chi.
\label{PR_PPR_free}
\end{equation}

Below, the expression for the limiting period $P_E$ has the standard form \cite{LPP1996}
\begin{eqnarray}
P_{E} \approx \frac{R}{c} \, \left(\frac{Rc^2}{GM}\right)^{1/2}
\, \frac{v_{\infty}}{c} \,
\left(\frac{B_0^2}{4\pi \rho_{\infty}v_{\infty}^2}\right)^{1/4}
\times \nonumber \\ \times
\left(\frac{c_{\infty}}{v_{\infty}}\right)^{1/2}
\approx 10^2 \, \frac{\mu_{30}^{1/2} c_{7}^{1/2} v_{7}^{1/2}}
{(B_{\rm ext})_{-6}^{1/2}}\mbox{ s.} \hspace{1em}
\end{eqnarray}

\noindent Here, $\rho_{\infty}$ and $c_{\infty}$ are, respectively, the density and the speed of sound in the interstellar medium ($c_7$ and $v_7$ are in the units of $10^7 \mbox{ cm s}^{-1}$); $B_{ext}^2 = 8\pi \rho_{\infty} c_{\infty}^2$ is the corresponding magnetic energy density ($(B_{\rm ext})_{-6}$ is in units of $10^{-6}$ G); and $\mu_{30}$ is the magnetic moment of the neutron star in units of $10^{30} \mbox{ G cm}^3$. In addition, we assumed that the velocity of neutron stars $v_{\infty}$ is much larger than the speed of sound in the interstellar medium.

Next, it is necessary to determine the spin period at which the star passes to the subsonic propeller stage. For this purpose, we recall that a pulsar can be at the supersonic propeller stage as long as its angular velocity is large enough, and, accordingly, the centrifugal acceleration exceeds the free-fall acceleration. Consequently, the condition under which the accretion of matter from the interstellar medium onto an isolated neutron star begins can be specified as follows: 
\begin{equation}
\omega^2R_A=\frac{GM}{R_A^2},
\end{equation}

\noindent where $\omega=2 \pi/P$ is the angular velocity of the neutron star.

As a result, for the boundary period corresponding to the transition of an isolated neutron star to the subsonic propeller stage, we obtain an expression that is similar to that derived by Ikhsanov \cite{I2003}: 
\begin{equation}
P_{eq} \simeq 24 \mu_{30}^{6/7} \dot M_{15}^{-3/7} m ^{-5/7} \mbox{ s},
\end{equation}

\noindent where $\dot M_{15}$ is the accretion rate in  units of $10^{15} \mbox{ g s}^{-1}$, and $m=M/M_{\bigodot}$.

Substituting the parameters into the above formula, we can estimate the period at which the neutron star leaves the supersonic propeller stage: 
\begin{equation}
P_{eq} \simeq 670 B_{12}^{6/7} \mbox{ s}.
\label{PR_Peq}
\end{equation}

Let us now consider the spin period distribution of neutron stars passed to the supersonic propeller stage for the standard model, i.e., for the model in which the evolution of the inclination angle in the region of extinct radio pulsars is disregarded. In this case, the kinetic equation can be written as
\begin{equation}
\frac{d}{dP} \left(N_3 \frac{dP}{dt} \right)=0.
\end{equation}

It should also be recalled that $N_2(P)=N_f P$ for the region of extinct radio pulsars in the standard model, while the derivative of the spin period at the supersonic propeller stage can be represented by Eq.~(\ref{PR_dP_dt}). As a result, we obtain the following distribution of neutron stars in spin period $P$ for the model where the evolution of the angle of inclination of the magnetic axis to the spin axis is disregarded:
\begin{eqnarray}
N_3(P)=1.4 \cdot 10^{-2} N_f P^{-2}, \mbox{  if  }P>P_E,
\nonumber \\
N_3(P)=0, \hspace{6em} \mbox{  if  }P<P_E,
\end{eqnarray}

In this case, the total number of neutron stars at the supersonic propeller stage for the model in which the inclination angle is disregarded can be estimated as $N_3=1.4 \times 10^{-2} N_f/P_E$.

\section{Model with hindered particle escape from the neutron-star surface}

Let us consider the statistics of pulsars at the supersonic propeller stage in more detail assuming that the model with hindered particle escape from the neutron-star surface is valid. In this case, the boundary of the transition from the region of extinct radio pulsars (the ejector stage) to the supersonic propeller region is defined by Eq. (\ref{PR_PPR_hind}).

Thus, the source function $F(P,B)$ in the kinetic equation (\ref{PR_kin}) can be written as 
\begin{equation}
F(P,B)=F(P,B, \chi)|_{sin\chi=P^2P_E^{-2}},
\nonumber
\end{equation}

\noindent where 
\begin{equation}
F(P,B,\chi)=\dot \chi N_2.
\nonumber
\end{equation}

The distribution function $N_2(P,B, \chi)$ in the latter expression refers to the region of extinct radio pulsars, and we found it previously \cite{BE2005}. Recall that this function for the model with hindered particle escape from the neutron-star surface can be specified as
\begin{equation}
N_2[P, B, \chi]= 0.7 k_N N_f B_{12}^{8/7}(1+B_{12})^{-3.7} \frac{G(cos \chi/P)}{sin \chi},
\nonumber
\end{equation}

\noindent where
\begin{equation}
G(\xi)=\frac{1-c(\xi)}{s^3(\xi)}(3.5+\frac{s^2(\xi)}{c^2(\xi)})c^{1.7}(\xi).
\nonumber
\end{equation}

\noindent In this case, we used two auxiliary functions:
\begin{eqnarray}
c(\xi)=B_{12}^{0.75} \xi^{1.4}=cos \chi,
\nonumber
\\
s(\xi)=(1-c^2(\xi))^{1/2}=sin \chi.
\nonumber
\end{eqnarray}

To determine the derivative of the angle of inclination of the magnetic axis to the spin axis of the neutron star, we use the integral of motion in explicit form. If the Ruderman-Sutherland model is valid, then the plasma in neutron stars at the stage of extinct radio pulsar will fill only the inner regions of the magnetosphere. Consequently, the energy losses for such pulsars can be assumed to be identical to the magnetosphere losses. Thus, the quantity $I=cos \chi /P$ is conserved during the evolution. As a result, we obtain the derivative of the angle of axial inclination in the region of extinct radio pulsars
\begin{equation}
\dot \chi=-\frac{\dot P}{P}cot \chi.
\nonumber
\end{equation}

The differential equation for the distribution function $N_3(P,B)$ of the pulsars at the supersonic propeller stage then takes the form

\begin{equation}
\frac{d}{dP} \left(N_3 P^{-2}\right)=0.01 k_N N_f B_{12}^{20/7}(1+B_{12})^{-3.7} \frac{y}{P^2}G\left(\frac{y}{P}\right).
\nonumber
\end{equation}

\noindent Here, we denoted $y=(1-P^4 P_E^{-4})^{1/2}$.

The sought-for distribution function can then be represented as
\begin{equation}
N_3(P,B)=0.01 k_N N_f B_{12}^{20/7}(1+B_{12})^{-3.7}P^{-2}
\int \limits_{P_{min}}^P \frac{y}{P^2}G\left(\frac{y}{P}\right)dP,
\nonumber
\end{equation}

\noindent where $k_N \approx 4.4$ is the normalization factor, and $N_f$ is the number of normal radio pulsars far from the death line (for more detail, see \cite{BE2005}).

Since the death line of normal radio pulsars 
\begin{equation}
P_d(\chi)=B_{12}^{8/15}(cos \chi)^{0.29} \mbox{ s}
\label{PR_deadline}
\end{equation}

\noindent crosses the boundary of the transition to the supersonic propeller stage at a fairly small angle $\chi$ ($sin \chi=P_E^{-2}P^{0.125}$), the point of intersection of the death line for normal radio pulsars (\ref{PR_deadline}) with the $sin \chi=0$ axis can be taken with a good accuracy as the point with the minimum period $P_{min}$ (see Fig.~\ref{fig1}). It thus follows that 
\begin{equation}
P_{min}=B_{12}^{8/15} \mbox{ s}.
\nonumber
\end{equation}

Since the period at which a pulsar passes from the ejector stage to the propeller stage is defined by Eq. (\ref{PR_PPR_hind}) and the period at which a pulsar leaves the supersonic propeller stage is defined by Eq. (\ref{PR_Peq}), we derive the following expression for the period distribution function of neutron stars at the supersonic propeller stage: 
\begin{eqnarray}
N_3(P,B)=0.01 k_N N_f P^{-2} 
\int \limits_{5 \cdot 10^{-4} P^{7/6}}^{P^{15/8}}
B_{12}^{20/7}(1+B_{12})^{-3.7}\times
\nonumber \\ \times
\int \limits_{B_{12}^{8/15}}^P
\frac{y}{P^2} G \left( \frac{y}{P} \right) dP. \hspace{12em}
\label{}
\end{eqnarray}

\renewcommand{\baselinestretch}{1.0}

\begin{figure}
\centering
\includegraphics[width=10cm,height=10cm,keepaspectratio]{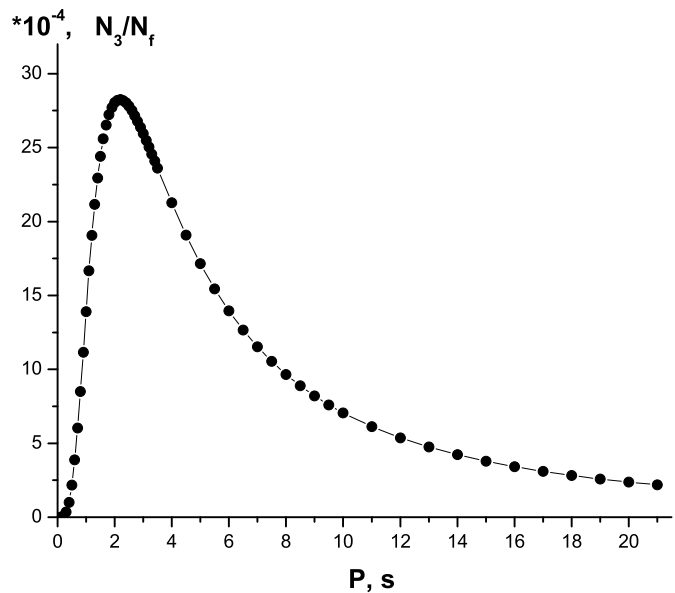}
\caption{\small{Distribution of neutron stars at the supersonic propeller stage $N_3(P)$ for the model with hindered particle escape.}}
\label{fig3}
\end{figure}

\renewcommand{\baselinestretch}{1.5}

Figure~\ref{fig3} shows the distribution of pulsars at the supersonic propeller stage in period $P$ for three different values of $P_E$: 10, 30, and 100 s. This distribution was constructed by assuming that the model with hindered particle escape from the neutron-star surface was valid. In addition, we assumed that no neutron stars are born at the supersonic propeller stage and did not consider the possibility of magnetic-field evolution over the lifetime of the pulsar at this stage. As we see from this figure, the number of such neutron stars is much smaller than the number of both active and extinct radio pulsars.

As was noted above, when allowance is made for the evolution of the angle of inclination of the magnetic axis to the spin axis in the region of extinct radio pulsars (and, accordingly, for the boundary of the transition to the supersonic propeller stage), the transition to the propeller stage can occur even at fairly short spin periods of the neutron star. The numerically computed distribution of pulsars at the supersonic propeller stage shown in the figure confirms this conclusion. We also see from Fig.~\ref{fig3} that the total number of neutron stars increases compared to the standard model when allowance is made for the evolution of the inclination angle in the region of extinct radio pulsars and, accordingly, for the boundary of the transition to the propeller stage.

It should also be noted that if the model with hindered particle escape is valid, the limiting period $P_E$ affects only slightly the distribution of neutron stars at the supersonic propeller stage. This situation arises from the fact that most of the neutron stars crossing the boundary of the region of extinct radio pulsars $sin \chi=P^2/P_E^2$ have short spin periods, $P \ll 100$ s (Fig.~\ref{fig1}).

\section{Model with free particle escape from the neutron-star surface}

Recall that the boundary of the transition from the region of extinct radio pulsars to the supersonic propeller stage for the model with free particle escape from the neutron-star surface is specified by Eq. (\ref{PR_PPR_free}) (see Fig.~\ref{fig2}). Consequently, the following relation is valid for the source function $F(P,B)$ determined by the neutron-star flux from the region of extinct radio pulsars: 
\begin{equation}
F(P,B)=F(P,B,\chi)|_{cos\chi=P^2 P_E^{-2}}.
\label{PR_F_PB}
\end{equation}

\noindent Therefore, as for the Ruderman-Sutherland model, the source function $F(P,B)$ can be written as
\begin{equation}
F(P,B,\chi)=\dot \chi N_2.
\label{PR_F_PBchi}
\end{equation}

If the Arons's model is valid for the existing neutron stars, then, as was mentioned above, the neutron-star magnetosphere for extinct radio pulsars still remains completely filled with plasma, which screens the magnetic field. As a result, all of the energy losses are attributable to longitudinal currents, while the magnetodipole losses are completely screened. In this case, the quantity $I_d=\sin \chi/P$ is conserved during the evolution at the ejector stage (including the region of extinct radio pulsars). It thus follows that the derivative of the angle of inclination of the magnetic axis to the spin axis of the neutron star is
\begin{equation}
\dot \chi=-\frac{\dot P}{P}tan \chi.
\label{PR_dot_chi}
\end{equation}

Based on our previous results, we use the following expression for the distribution function of extinct radio pulsars 
\begin{equation}
N_2(P,B,\chi)=0.3 k_N N_f \frac{B_{12}^{1.1}\left(1+B_{12}\right)^{-3.7}}
{cos \chi \left(1+0.1B_{12}^{1.7}\left(sin \chi/P\right)^{3.1}\right)^{0.9}}.
\label{PR_N2}
\end{equation}

Let us change the variable
\begin{equation}
y=sin \chi=1-\frac{P^4}{10^8B_{12}^{1/2}}.
\nonumber
\end{equation}

\noindent Substituting (\ref{PR_F_PB}), (\ref{PR_F_PBchi}), (\ref{PR_dot_chi}), and (\ref{PR_N2}) into the kinetic equation (\ref{PR_kin}) yields the differential equation for $N_3(P,B)$
\begin{eqnarray}
\frac{d}{dp}(N_3 P^2)= 4 \times 10^{-3}P_E k_N N_f B_{12}^2.74 (1+B_{12})^{-3.7} P^{-27/14}
\times \nonumber \\ \times
\frac{y}{\left(1+0.1B_{12}^{1.7}\left(y/P \right)^{3.1}\right)^{0.9}}.
\nonumber
\end{eqnarray}

The solution of this equation is the sought-for distribution function
\begin{eqnarray}
N_3(P,B)=4 \times 10^{-3}P_E k_N N_f B_{12}^{2.74}(1+B_{12})^{-3.7}P^{-2}
\times \nonumber \\ \times
\int \limits_{P_{min}}^P \frac{y}{P^{27/14}} \frac{dP}{\left(1+0.1B_{12}^{1.7} \left(y/P\right)^{3.1} \right)^{0.9}}.
\label{PR_N3}
\end{eqnarray}

To find minimum spin period of a neutron star that passes to the supersonic propeller stage, let us turn to Fig.~\ref{fig2}. We see from this figure that the point of intersection between the line of transition of pulsars to the region where the angle of inclination of the magnetic axis to the spin axis is close to $90^{\circ}$ ($cos \chi=\left(\Omega R/c \right)^{1/2}$) and the line of transition from the region of extinct radio pulsars to the propeller stage (\ref{PR_PPR_free}), where $\Omega=2 \pi/P$ is an angular velocity, corresponds to this period. Thus, the minimum spin period of a neutron star at the supersonic propeller stage is
\begin{equation}
P_{min}=0.184 A^{4/5} B_{12}^{2/5}.
\nonumber
\end{equation}

\noindent Here, we represented the limiting period $P_E$ as $P_E=AB_{12}^{1/2}$.

As a result, we obtain the spin period distribution of neutron stars at the supersonic propeller stage for the model with free particle escape from the pulsar's surface by taking into account the boundaries of the transition of the neutron star to the supersonic propeller stage (\ref{PR_PPR_free}) and the subsonic propeller stage (\ref{PR_Peq}) and using the distribution function (\ref{PR_N3}):
\begin{eqnarray}
N_3(P)= 4 \times 10^{-3} P_E k_N N_f P^{-2} \int \limits_{5 \times 10^{-4}P^{7/6}}^{69 \cdot A^{-2}P^{5/2}} B_{12}^{2.74}(1+B_{12})^{-3.7}
\times \nonumber \\ \times
\int \limits_{0.185P_E^{0.8}}^{P} \frac{y}{P^{27/14}}
\frac{dP}{\left(1+0.1B_{12}^{1.7}(y/P)^{3.1} \right)^{0.9}}.
\nonumber
\end{eqnarray}

\noindent Recall that in this expression, the normalization factor is $k_N \approx 4.4$, and $N_f$ is the number of normal radio pulsars far from the death line.
\renewcommand{\baselinestretch}{1.0}

\begin{figure}
\centering
\includegraphics[width=10cm,height=10cm,keepaspectratio]{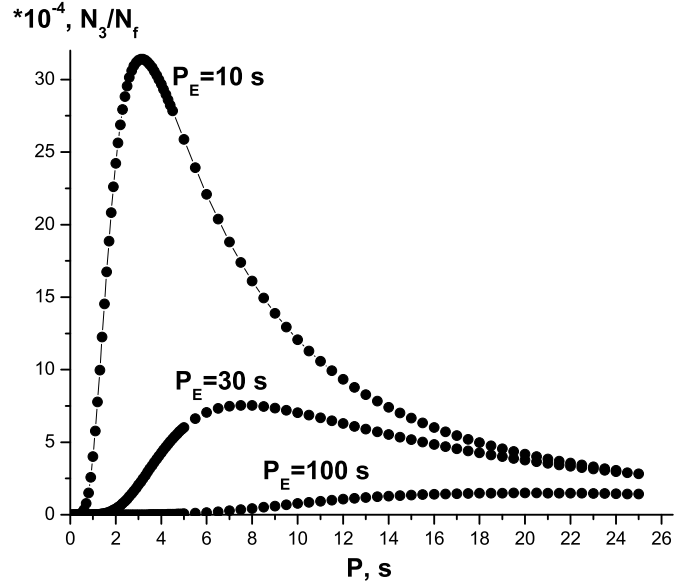}
\caption{\small{Distribution of neutron stars at the supersonic propeller stage $N_3(P)$ for the model with free particle escape.}}
\label{fig4}
\end{figure}

\begin{figure}
\centering
\includegraphics[width=10cm,height=10cm,keepaspectratio]{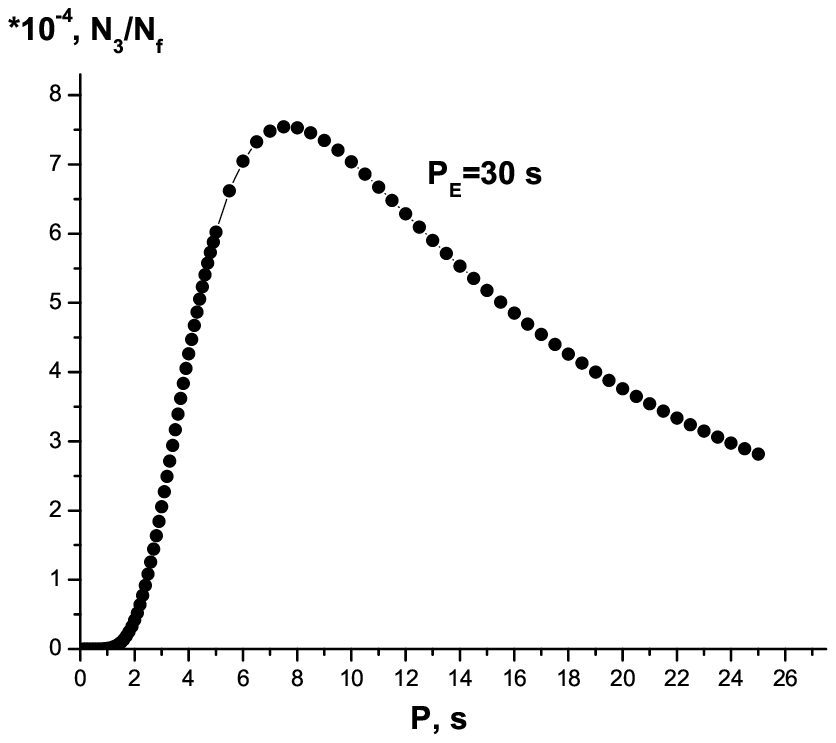}
\caption{\small{Distribution of neutron stars at the supersonic propeller stage $N_3(P)$ for the model with free particle escape for the limiting period $P_E=30$ s.}}
\label{fig5}
\end{figure}

\begin{figure}
\centering
\includegraphics[width=10cm,height=10cm,keepaspectratio]{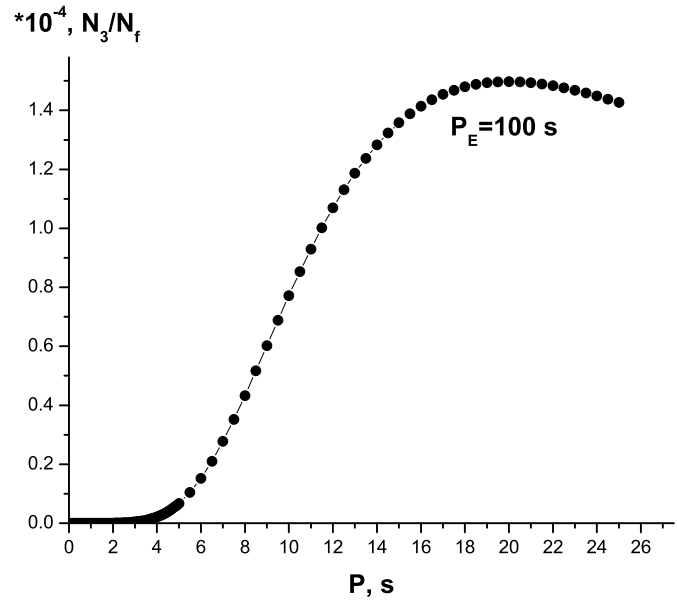}
\caption{\small{Same as Fig.~\ref{fig5} for $P_E=100$ s.}}
\label{fig6}
\end{figure}

\renewcommand{\baselinestretch}{1.5}

Figure~\ref{fig4} shows the distribution of neutron stars at the supersonic propeller stage in spin period $P$ under the assumption that the Arons's model is valid for three limiting periods $P_E$: 10, 30, and 100 s. In addition, Figs.~\ref{fig5} and \ref{fig6} separately show the distributions for $P_E=30 \mbox{~s}$ and $P_E=100 \mbox{~s}$, respectively. As one can see from these figures, when allowance is made for the evolution of the inclination angle $\chi$ at the ejector stage and, accordingly, for the boundary of the transition to the propeller stage, the number of neutron stars at the supersonic propeller stage is indeed found to be significant at fairly short spin periods ($P \sim 5-10$ s). Thus, a consistent allowance for the angle of inclination of the magnetic axis to the spin axis of the neutron star at the ejector stage leads to an increase in the number of pulsars that passed to the supersonic propeller stage.

It should also noted that, in contrast to the model with hindered particle escape from the neutron-star surface, the limiting period $P_E$ plays a significant role in the distribution of neutron stars at the supersonic propeller stage for the Arons's model. This is because the distribution of neutron stars at the boundary of the transition from the region of extinct radio pulsars to the propeller stage is virtually uniform for this model.

\section{Conclusions}

We have determined the distributions of neutron stars at the supersonic propeller stage in spin period $P$ taking into account the fact that a consistent allowance for the evolution of the angle of inclination of the magnetic axis to the spn axis for the region of extinct radio pulsars directly affects the statistics of neutron stars at the supersonic propeller stage. All our calculations were performed for two models of the particle acceleration region: the model with hindered particle escape \cite{RS1975} and the model with free particle escape \cite{A1979}.

As a result we have made sure once again that a consistent allowance for the evolution of the inclination angle $\chi$ in the region of extinct radio pulsars both for the model with hindered particle escape and for the model with free particle escape leads to an increase in the total number of neutron stars at the supersonic propeller stage. This increase stems from the fact that when the inclination angle $\chi$ for the evolution of neutron stars in the region of extinct radio pulsars and, hence, for the boundary of the transition to the supersonic propeller stage is included, this transition is possible at shorter spin periods of the neutron star ($P \sim 5-10$~s) than assumed in the standard model.

\section{Acknowledgements}

We would like to thank S.B.~Popov and M.E.~Prokhorov for fruitful discussions and numerous advice. This work was supported by the Russian Foundation for Basic Research (project no.05-02-17700).

\end{document}